\documentclass[twocolumn]{aastex63}
\usepackage{graphicx}
\usepackage{blindtext}
\usepackage{amsmath}
\usepackage{mathtools}
\usepackage{multirow}
\usepackage{hyperref}
\hypersetup{
	colorlinks	= true,
	linkcolor	= red,
	urlcolor	= cyan,
	citecolor	= blue
}
\usepackage{longtable}
\usepackage{lineno}
\usepackage{array}
\newcolumntype{C}[1]{>{\centering\arraybackslash}p{#1}}
\usepackage{makecell}

\newcommand{\exoclock}{\mbox{ExoClock}}

\begin{document}

\title{ExoClock Project III: 450 new exoplanet ephemerides from ground and space observations}

\correspondingauthor{A. Kokori}

\email{anastasia.kokori.19@ucl.ac.uk}

\author{A. Kokori}
\affiliation{University College London, Gower Street, London, WC1E 6BT, UK}

\author{A. Tsiaras}
\affiliation{INAF - Osservatorio Astrofisico di Arcetri, Largo E. Fermi 5, 50125 Firenze, Italy}\affiliation{University College London, Gower Street, London, WC1E 6BT, UK}

\author{B. Edwards}
\affiliation{AIM, CEA, CNRS, Université Paris-Saclay, Université de Paris, F-91191 Gif-sur-Yvette, France}
\affiliation{University College London, Gower Street, London, WC1E 6BT, UK}

\author{A. Jones}
\affiliation{Amateur Astronomer\footnote{A list of associated private observatories that contributed to this work can be found in Appendix A}}\affiliation{British Astronomical Association, Burlington House, Piccadilly, Mayfair, London, W1J 0DU, UK}

\author{G. Pantelidou}
\affiliation{Department of Physics, Aristotle University of Thessaloniki, University Campus, Thessaloniki, 54124, Greece}

\author{G. Tinetti}
\affiliation{University College London, Gower Street, London, WC1E 6BT, UK}

\author{L. Bewersdorff}
\affiliation{Amateur Astronomer\footnote{A list of associated private observatories that contributed to this work can be found in Appendix A}}

\author{A. Iliadou}
\affiliation{Department of Physics, Aristotle University of Thessaloniki, University Campus, Thessaloniki, 54124, Greece}

\author{Y. Jongen}
\affiliation{Amateur Astronomer\footnote{A list of associated private observatories that contributed to this work can be found in Appendix A}}\affiliation{Observatoire de Vaison-La-Romaine, Départementale 51, près du Centre Equestre au Palis - 84110 Vaison-La-Romaine, France}

\author{G. Lekkas}
\affiliation{Department of Physics, University of Ioannina, Ioannina, 45110,  Greece}

% ...

\begin{abstract}

The \exoclock\ project has been created with the aim of increasing the efficiency of the Ariel mission. It will achieve this by continuously monitoring and updating the ephemerides of Ariel candidates over an extended period, in order to produce a consistent catalogue of reliable and precise ephemerides. This work presents a homogenous catalogue of updated ephemerides for 450 planets, generated by the integration of $\sim$18000 data points from multiple sources. These sources include observations from ground-based telescopes (\exoclock\ network and ETD), mid-time values from the literature and light-curves from space telescopes (Kepler/K2 and TESS). With all the above, we manage to collect observations for half of the post-discovery years (median), with data that have a median uncertainty less than one minute. In comparison with literature, the ephemerides generated by the project are more precise and less biased. More than 40\% of the initial literature ephemerides had to be updated to reach the goals of the project, as they were either of low precision or drifting. Moreover, the integrated approach of the project enables both the monitoring of the majority of the Ariel candidates (95\%), and also the identification of missing data. The dedicated \exoclock\ network effectively supports this task by contributing additional observations when a gap in the data is identified. These results highlight the need for continuous monitoring to increase the observing coverage of the candidate planets. Finally, the extended observing coverage of planets allows us to detect trends (TTVs - Transit Timing Variations) for a sample of 19 planets. All products, data, and codes used in  this work are open and accessible to the wider scientific community. 
\end{abstract}


\section{INTRODUCTION}

The number of exoplanets discovered already exceeds 5000 and it continues to increase daily. The characterisation of exoplanets will be the main goal for future space missions. The Ariel mission aims to observe the atmospheres of 1000 planets in 2029 in order to investigate their nature \citep{2018ExA....46..135T}. Ariel will observe thousands of transits, and to increase the mission efficiency, it is required to have precise ephemerides. Proper planning is important to avoid wasting the precious observing time of Ariel and other future space missions. 

For various reasons, accuracy in predicting transit times is impeded. For example, the uncertainties of the initial ephemerides causes degeneracies over time in the precision of the predicted transit time \citep[e.g.][]{2019AnA...622A..81M}. The insufficient number of available data for each planet is another factor that generates biases in calculating the ephemerides \citep[e.g.][]{2017ApJ...834..187B, 2019AnA...622A..81M}. 

To overcome the above problems and create a complete catalogue of precise ephemerides for a large number of planets, it is essential to use all available resources of data. These resources consist of data from the literature, data obtained by telescopes from the ground and finally data from space telescopes. The \exoclock\ project \citep{Kokori2021,2022ApJS..258...40K} is an open, integrated platform, with the aim of continuously monitoring the ephemerides of the Ariel candidate targets \citep{Edwards2022}. The organisation of the project is described thoroughly in \cite{Kokori2021}, and the first large-scale catalogue of updated ephemerides for 180 planets was produced in \cite{2022ApJS..258...40K}, by combining observations from ground-based telescopes and literature. 

The benefits of using small, ground-based telescopes to observe transiting exoplanets have been underlined previously \citep[e.g.][]{2020PASP..132e4401Z,2019AnA...622A..81M,2021MNRAS.504.5671E,2019PASP..131h4402B}, and their use for large-scale studies has been proved already in \cite{2022ApJS..258...40K}. While small telescopes are efficient, maximum effectiveness is achieved by utilising all available resources including other ground based networks, data from the literature and also data from space resources. Space telescopes are effective for observing challenging transits not easily accessible from ground telescopes. Moreover, TESS \citep{Ricker2014} has been scanning the sky since 2019 and will continue to provide light-curves for many known exoplanets, which can be used for ephemerides updates \citep{2022ApJS..259...62I}. Therefore, to produce a complete catalogue of ephemerides for all planets, it is important to use data from both space- and ground-based telescopes.

In this study, we integrated data from the \exoclock\ network, mid-time points from the literature, data from the Exoplanet Transit Database \citep{Poddany2010} and data from space telescopes (Kepler \citep{Koch2010}, K2 \citep{Howell2014}, and TESS \citep{Ricker2014}). The integration of $\sim$18000 mid-time points, in total, allowed the generation of a complete analysis of ephemerides for 450 planets. The benefits of this integrated analysis are several: biases are minimised, better precision is achieved and long-term phenomena can be identified for each planet which may be indicative of trends (e.g. TTVs). The integrated design and approach of the \exoclock\ project highlights where there are gaps in the available data. The \exoclock\ network of ground-based telescopes can then be directed flexibly to make observations to address such gaps and extend the coverage. 

The \exoclock\ project operates with an Open Science Framework in all the aspects of the research cycle \citep[\href{https://digital-strategy.ec.europa.eu/en/library/open-innovation-open-science-open-world?fbclid=IwAR2QVLOQCeb_vMvXltNXoD_iDLJho5r95NB_7mywVJKnN63hfd7fOuyOgT0}{EU 2016};][]{2dbff737-en}. Open Science advances the progress of scientific research by encouraging collaborations and reproducibility. During all stages of the scientific process, the project follows open science practices; data, tools and codes used for the analysis are all open and accessible to everyone.  All data used in this study are open and publicly available (e.g. data obtained from the \exoclock\ network, data from space telescopes). Additionally, open science means co-creation of scientific research through collaborations between various scientific communities and also citizen involvement (\href{https://digital-strategy.ec.europa.eu/en/library/open-innovation-open-science-open-world?fbclid=IwAR2QVLOQCeb_vMvXltNXoD_iDLJho5r95NB_7mywVJKnN63hfd7fOuyOgT0}{EU 2016}). In this respect, the project is open to contributions from any interested person or community. The collaborative perspective also helps to ensure the most effective use of resources. Such collaborations are important to avoid overlapping and waste of observing time and to a further extent, they foster innovation in science.

\section{DATA}

\begin{table*}
\centering
\caption{Summary of the observations used in this work. As coverage we define the percentage of years (since the first observation in the database) for which at least one observation exists.}
\label{tab:data}
\begin{tabular}{c c c c c c c c}
 							& ExoClock 	& ETD 		& Kepler 	& K2 		& TESS 		& Literature  	& \textbf{Total}		\\ \hline
Data points					& 2911		& 184		& 5763		& 371		& 6499		& 2442			& \textbf{18170}		\\ 
Years						& 2007-2021	& 2001-2021	& 2009-2013	& 2014-2018	& 2018-2021	& 2004-2020		& \textbf{2001-2021}	\\
Planets						& 302		& 40		& 21		& 49		& 371		& 340			& \textbf{450}			\\
Median $\sigma_{T_{mid}}$ 	& 1.3 min	& 1.7 min	& 0.5 min	& 0.6 min	& 1.1 min	& 0.6 min		& \textbf{0.8 min}		\\
Median coverage				& 20\%		& 18\%		& 29\%		& 17\%		& 20\%		& 17\%			& \textbf{50\%}			\\

\end{tabular}
\end{table*}

In this study, we integrated 
light-curves from the \exoclock\ network, the Exoplanet Transit Database, and the MAST Archive (for the Kepler, K2, and TESS space missions), 
and mid-transit times from the literature to update the ephemerides of 450 exoplanets. All the light-curves were acquired before the end of 2021 and the literature mid-transit times were published by the end of 2021. We analysed all the light-curves, regardless of their source, using the stellar and planetary parameters included in the Exoplanet Characterisation Catalogue (ECC), a dedicated catalogue prepared and maintained within the \exoclock\ project \citep{Kokori2021}, and the open-source Python package PyLightcurve \citep{Tsiaras2016B2016ascl.soft12018T}. For every light-curve, PyLightcurve peroforms the following operations:: 

\begin{figure}
\centering
\includegraphics[width=\columnwidth]{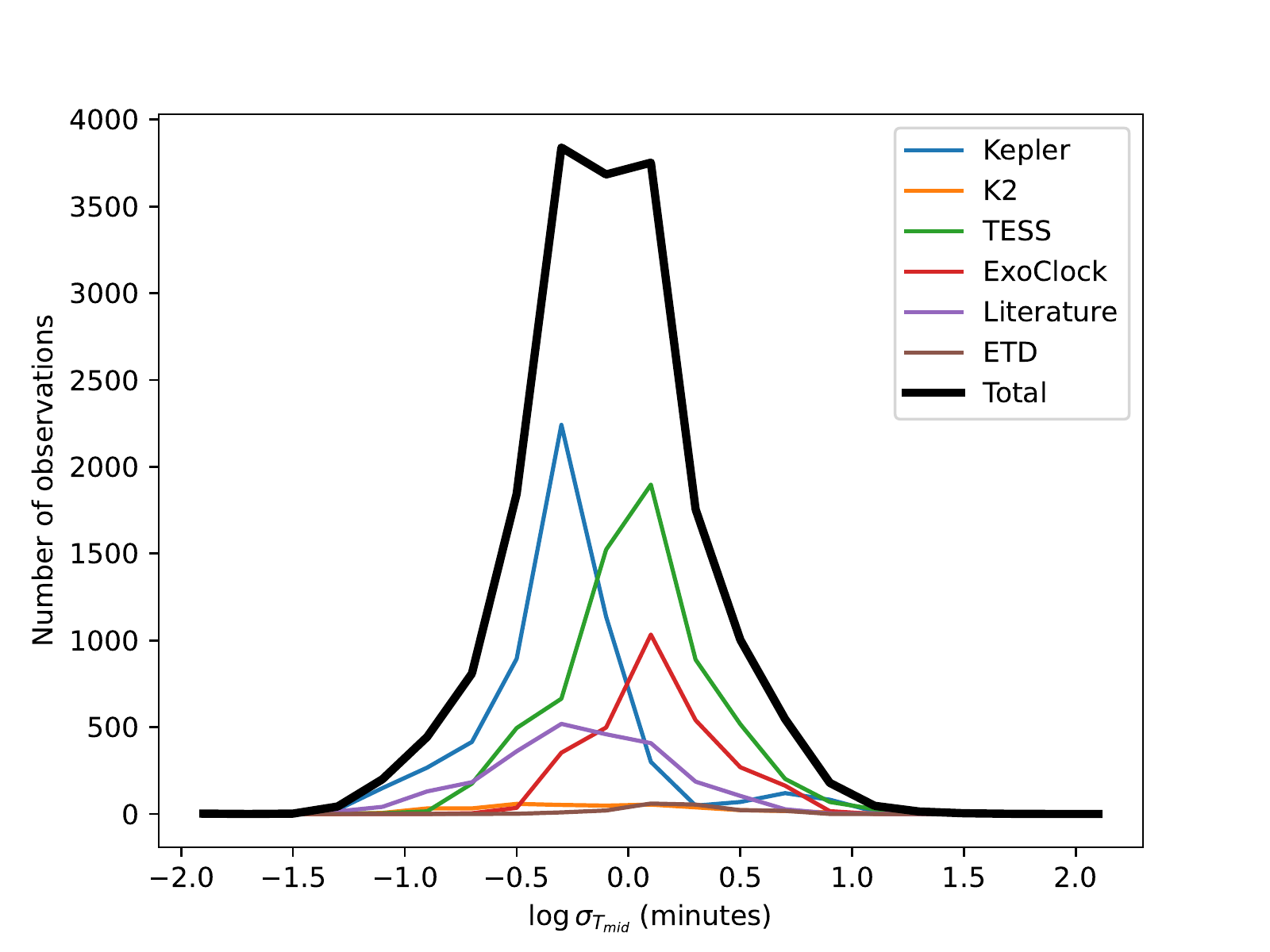}
\caption{Distribution of transit mid-time uncertainties among the different sources.}
\label{fig:precision}
\end{figure}

\begin{figure}
\centering
\includegraphics[width=\columnwidth]{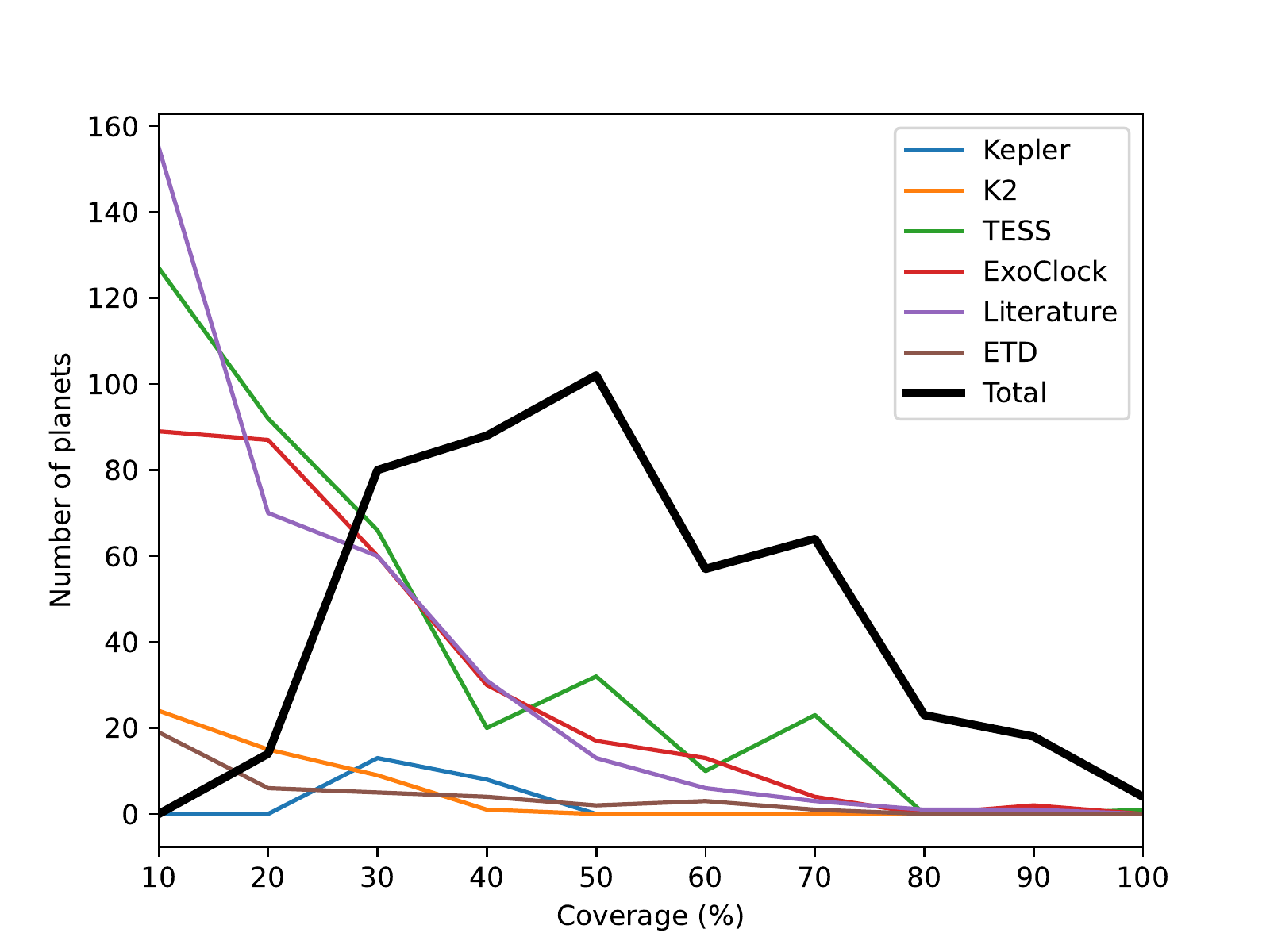}
\caption{Distribution of coverage among the different sources. As coverage we define the percentage of years (since the first observation in the database) for which at least one observation exists.}
\label{fig:coverage}
\end{figure}

\begin{enumerate}

\item calculates the limb-darkening coefficients   using the ExoTETHyS package \citep{Morello2020} (depending on the filter used for the observation),

\item converts the time formats to Barycentric Julian Date (BJD$_\mathrm{TDB}$),

\item finds the maximum-likelihood model for the data (an exposure-integrated transit model together with a trend model -- linear with airmass, linear with time, or quadratic) using the Nelder-Mead minimisation algorithm included in the SciPy package \citep{Virtanen2020},

\item removes outliers that deviate from the maximum-likelihood model by more than three times the standard deviation (STD) of the normalised residuals,

\item scales the uncertainties by the root mean square (RMS) of the normalised residuals, to take into account any extra scatter,

\item and, finally, performs an Markov chain Monte Carlo (MCMC) optimisation process using the emcee package \citep{ForemanMackey2013}

\end{enumerate}

After this analysis, the quality of each light-curve was evaluated individually, and light-curves that fulfilled one or more of the criteria below, were excluded:
     
\begin{enumerate}

\item autocorrelation and shapiro statistic indicate non-gaussian normalised residuals at a 3-$\sigma$ level or more,

\item transit signal-to-noise ratio (S/N $= Depth/\sigma_{Depth}$) is lower than three,

\item $Rp/Rs$ differs by more than 3$\sigma$ from the literature value (for the ExoClock and ETD observations), or the weighted average of the mission (for  the space observations),

\item O-C value is not in agreement (3$\sigma$) with other observations obtained during the same observing period ($\sim$a month).

\end{enumerate}

The final list of 450 planets includes those planets for which we collected data-points at three or more different epochs and we could determine an ephemeris of better or equal quality to the initial ephemeris. Table \ref{tab:data} summarises the observations used to produce the ephemerides of 450 planets in this work. In addition, Figures \ref{fig:precision} and \ref{fig:coverage} show the distribution of the precision and the coverage of the transit mid-time points used. As coverage we define the percentage of years (since the first observation in the database) for which at least one observation exists. We need to note here that 99\% of the observations used have transit mid-time uncertainties lower than 10 minutes, and that the median coverage of all sources combined together is 50\%, while individual sources do not reach more than 29\%.

\subsection{ExoClock - Summary and quality of data}

\begin{figure}
\centering
\includegraphics[width=\columnwidth]{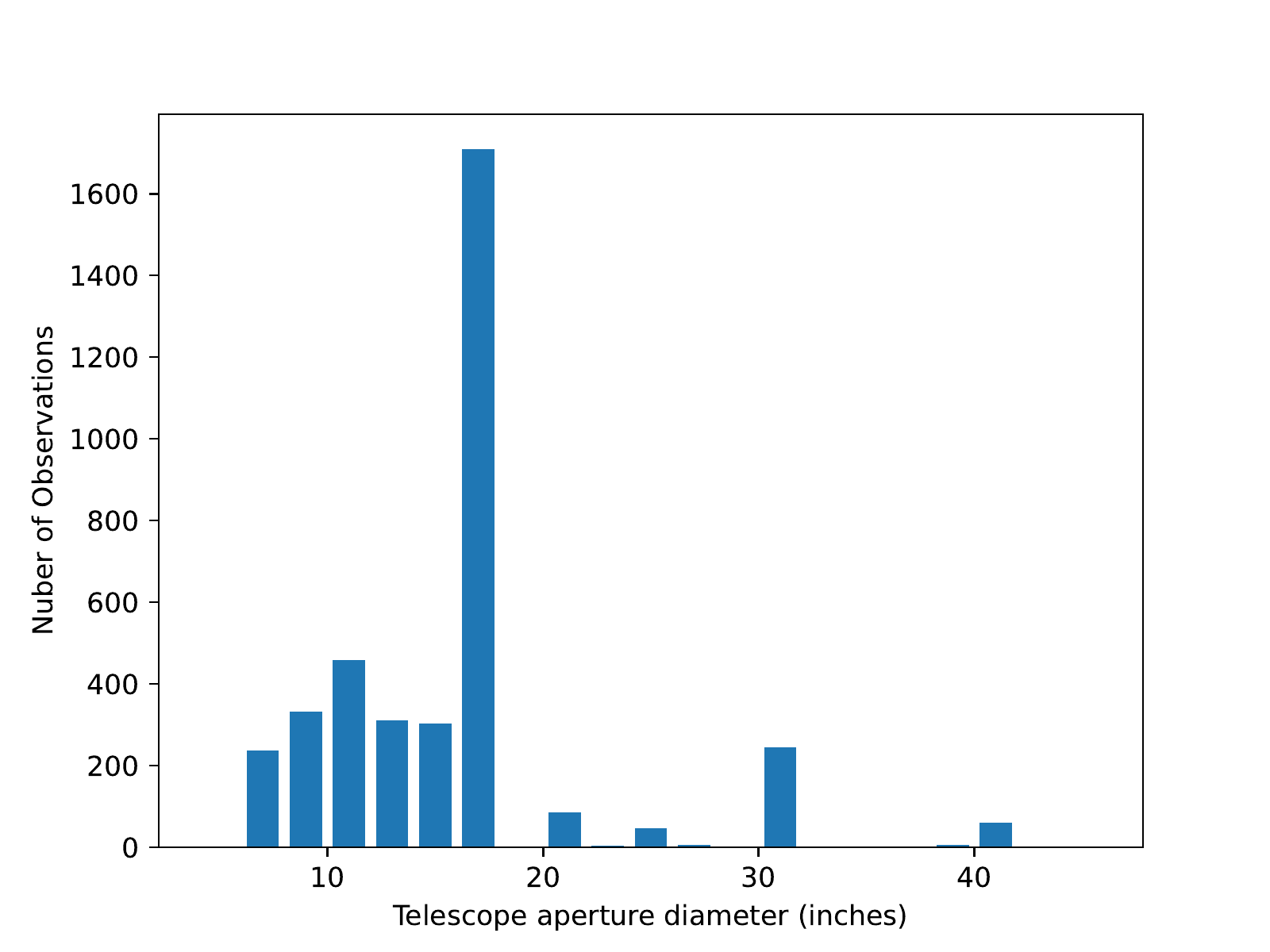}
\caption{Number of observations received from the \exoclock\ network, as a function of the telescope size.}
\label{fig:obs_vs_d}
\end{figure}

Currently, the \exoclock\ network consists of 540 participants -- 80\% of whom are amateur astronomers -- and 450 telescopes with sizes ranging between 6 and 40 inches -- of which 80\% are smaller than 17 inches. Figure \ref{fig:obs_vs_d} shows the distribution of the observations used in this work among the different telescope sizes. The large majority of the observations comes from small- and medium-scale telescopes and amateur observers (73\%), who are the key part of our network. The \exoclock\ network is organised in a way to maximise the coverage of the planets and to ensure the high quality and homogeneity of the results. To achieve this, we have defined a prioritisation system, we provide a personal scheduler for each telescope, we support the observers with the data analysis (educational material, a user-friendly software, regular meetings) and, finally, we perform the light curve modelling and evaluation (as described above) on the \exoclock\ website. For more details on the organisation of the \exoclock\ project and the \exoclock\ network, we refer the interested reader to \cite{Kokori2021}.

\subsection{Data from space telescopes}

For the first time in the \exoclock\ project, we integrated light-curves from space telescope observations. More specifically we included light-curves from Kepler, K2 and TESS (before the end of 2021). First, we downloaded the long-cadence light-curves for the targets in the \exoclock\ target list. Then we identified the transits inside those light-curves and isolated them, including a base-line of one transit duration before and one after the event. Finally, the analysis and evaluation of each light-curve was conducted as described above, using a quadratic de-trending function. As some of the space-based light-curves contained gaps, we only considered those light-curves that were at least 80\% complete, both in-transit and out-of-transit -- i.e total exposure time more then 0.8 times the transit duration before, during and after the transit.

From the analysis of the space-based light-curves, and especially from the TESS light-curves, it became clear that for a, non-negligible, number of planets, the parameters in the ECC (as derived from the literature) were producing transits of shorter or longer duration than the actual observations. For these planets we let the reduced semi-major axis ($a/R_s$) vary in order to account for the differences in the duration. The ECC has been updated accordingly and the planets for which the $a/R_s$ was adjusted have been marked. Table \ref{tab:parameters} includes the adjusted $a/R_s$ values, which are marked with an asterisk. With the exception of Kepler-396c, Kepler-854b, and TOI-201b, which had not value for their inclinations ($i$) in the discovery papers, we decided to fix $i$ to the literature value, as $a/R_s$ and $i$ are strongly correlated when they are both free parameters. In a future work we plan to provide analysis for both parameters but the scope of this work is to provide a set of parameters that will produce a reliable duration. Between $a/R_s$ and $i$ we decided to let $a/R_s$ free as it is a more flexible for the determination of the duration (only small changes are required, and there is no upper limit like $i$).

Finally, we need to note here that the modelling of the Kepler light-curves did not produce gaussian residuals. This is most probably due to the fixed limb-darkening coefficients (LDCs) used. However, we decided to not allow the LDCs to vary, in order to keep a homogeneous analysis pattern for all observations.

\begin{figure*}
\centering
\includegraphics[width=0.8\textwidth]{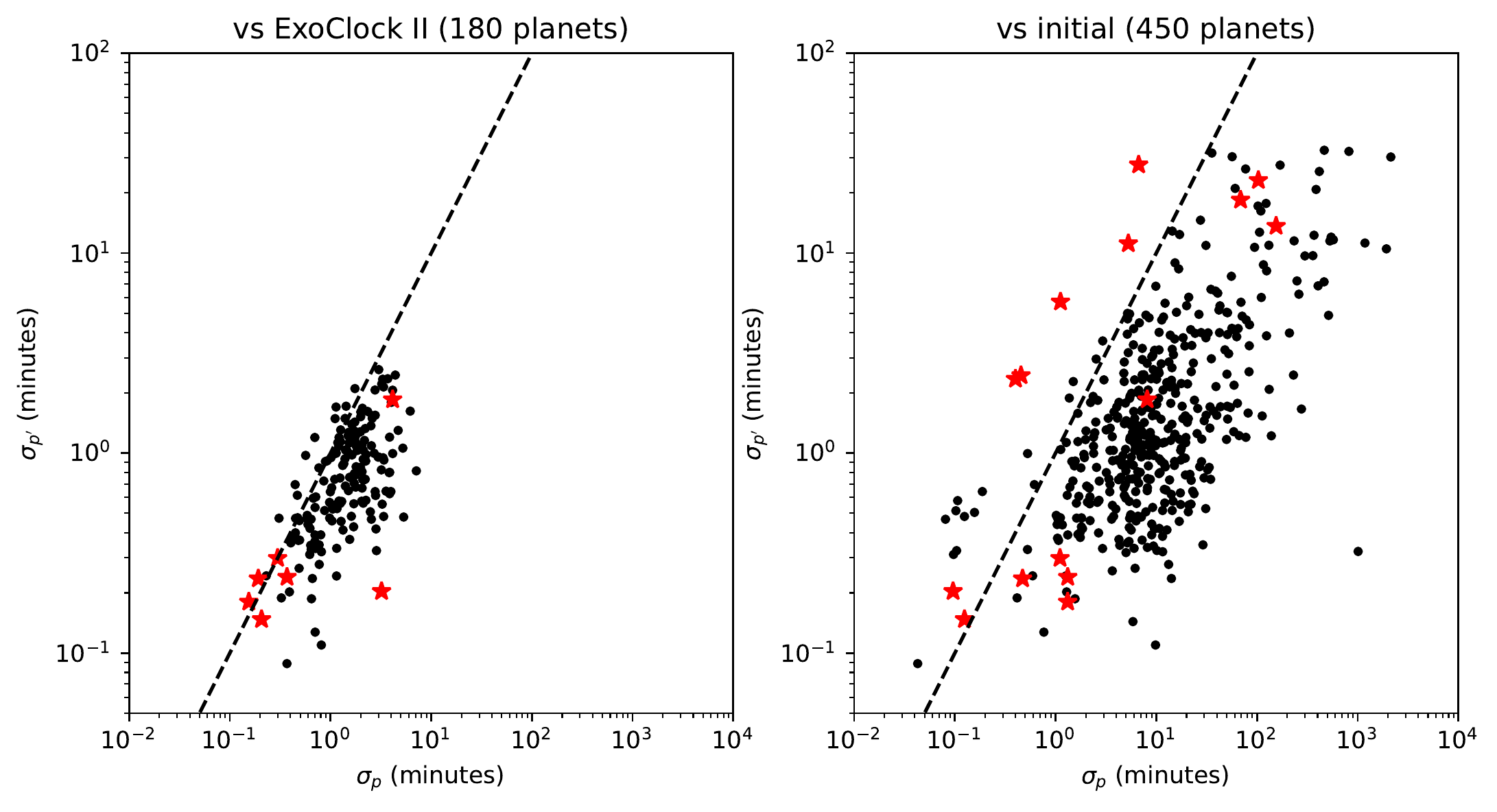}
\caption{Comparison of the 2029-prediction uncertainties between  this work and \exoclock\ II (left) and the ephemerides used at the beginning of the project (right). With the red star we indicate the planets for which TTV signals have been found. In both panels, the dashed lines are the $\sigma_{p'}=\sigma_{p}$ lines.}
\label{fig:before_after}
\end{figure*}

\subsection{Exoplanet Transit Database (ETD)}

The Exoplanet Transit Database \citep[ETD,][]{Poddany2010} run by the Czech Astronomical Society since 2009, is currently the largest database of transit follow-up observations with more than 10,000 transit light-curves for more than 350 exoplanetary systems. The collaboration between \exoclock\ and ETD started in 2020 and is described in \cite{2022ApJS..258...40K}. In this study, we included 184 observations for 40 planets provided by the ETD network. In order to maintain homogeneity and reliability in our analysis, the ETD observations were analysed and evaluated through the \exoclock\ website using the same methodology and validation criteria as for the \exoclock\ network data. The collaboration with ETD is critical to avoid duplications and waste of resources. We aim to continue our collaboration and gradually integrate more data from ETD in future publications. Such data can increase the coverage of certain planets during the period before \exoclock\ observations.

\subsection{Mid-time points from the literature}

As we did not reanalyse the original light-curves we could not apply the same criteria as for the \exoclock, ETD and space light-curves. From the available data we excluded mid-transit time values that referred only to ephemerides, rather than to individual transits (with the exception of the discovery papers). We also excluded mid-transit time values with uncertainties greater than five minutes, and mid-transit time values that originated from Kepler, K2, TESS or ETD, to avoid duplications.

\section{Results}

\begin{table}
\centering
\caption{Categories of ephemerides in comparison with the previous \exoclock\ publication and the values at the beginning of the project.}
\label{tab:before_after}
\begin{tabular}{c c c c c c c c}
 						& vs ExoClock II 	& vs initial	\\
   						& (180 planets) 	& (450 planets)	\\ \hline
Significantly improved	& 0.0\%				& 31.8\%		\\ 
Drifting				& 1.1\%				& 12.9\%		\\ 
Improved				& 31.7\%			& 41.1\%		\\ 
No change				& 63.3\%			& 10.0\%		\\
TTVs					& 3.9\%				& 4.2\%			\\
\end{tabular}
\end{table}

\subsection{Ephemerides}

Here we present updated ephemerides for 450 out of 570 planets that are currently in the \exoclock\ target list. To calculate the new ephemerides, we used all the available data from all the sources described in the previous section. First, we calculated an updated zero-epoch point as the weighted average of the available epochs. Then we fitted a line on the epoch vs mid-transit times data using the emcee package \citep{ForemanMackey2013}. After a first fit, we scaled-up the uncertainties by the RMS of the normalised residuals to account for excess noise, and performed the fit again. Table \ref{tab:updated_ephemerides} provides all the new ephemerides and references to the literature values used.

Figure \ref{fig:before_after} shows the uncertainties in the 2029-predictions before and after the updates presented in this work ($\sigma_p$ and $\sigma_{p'}$, respectively), while Table \ref{tab:before_after} lists five categories of the ephemerides status. ``Significantly improved'' refers to those ephemerides that were giving 2029-predictions with uncertainties greater than the target uncertainty of 1/12$^\mathrm{th}$ of the transit duration, D, ($\sigma_p>D/12$) as described in \cite{Kokori2021}. The term ``drifting'' refers to the ephemerides that were giving 2029-predictions that were drifting more than the target uncertainty ($|p-p'|>D/12$). From the remaining ephemerides, the term ``Improved'' refers to those ephemerides for which the 2029-prediction uncertainties have been improved by more than one minute ($\sigma_{p'}<\sigma_p - 1$), while ``No change'' refers to those ephemerides for which the 2029-prediction uncertainties have not changed by more than one minute ($|\sigma_{p'} - \sigma_p| < 1$). Finally, in this work we introduce the ``TTVs'' flag, which refers to ephemerides that deviate from a linear behaviour.

\subsection{Deviations from linear ephemerides}

For all the planets we calculated the Generalised Lomb-Scarge (GLS) periodogram on the linear ephemeris residuals to identify deviations from the linear ephemeris. We concluded that periodograms are more reliable in detecting such deviations, since other diagnostics such as the reduced chi square, the autocorrelation, or gaussianity tests on the residuals, are strongly affected by red noise, discontinuity and low number of data, respectively. This was due to the sparsity of the data and due to red noise in the timing measurements. The TTVs flag was given to those planets with periodograms that had peaks with a False Alarm Probability (FAP) lower than 0.13\%. We estimated the FAP for each planet as follows: first, we produced periodograms (Pa) for 100,000 series of white noise with the same sampling, then we produced periodograms (Pb) for 100,000 series where we varied the mid-time data within their uncertainties. Finally, the FAP for each period was defined as the percentage of Pb that had greater power than the 99.87\% (3$\sigma$) upper limit of the Pa periodograms. Detected periodicities were categorised as short-term or long-term, based on the time span of all available data. Long-term are these periodicities that are close to or longer than the total time span of the data used.

\section{Data release C}

The third data release of the \exoclock\ project includes two data products: the Catalogue of Observations (\exoclock, ETD, space observations), and the catalogue of \exoclock\ ephemerides. All data products and their descriptions can be found through the OSF repository with DOI: \href{http://doi.org/10.17605/OSF.IO/P298N}{10.17605/OSF.IO/P298N}.

\subsection{Catalogue of Observations}

The Catalogue of Observations contains all the light-curves and literature mid-time points summarised in Table \ref{tab:data}. In the online repository, each light-curve is accompanied by:
\begin{enumerate}
\item metadata regarding the planet, the source, the observation, the instrument, and the data format;
\item the pre-detrended light curve, filtered for outliers, converted to BJD$_\mathrm{TDB}$ and flux formats, with scaled uncertainties;
\item the fitting results, including the de-trending method used and its parameters;
\item the de-trended light curve, enhanced with the de-trending model, the transit model and the residuals;
\item fitting diagnostics on the residuals.
\end{enumerate}

\subsection{Catalogue of \exoclock\ Ephemerides} 

The new catalogue of \exoclock\ ephemerides contains the updated ephemerides for the 450 planets studied in this work (see also Table \ref{tab:updated_ephemerides}), accompanied by metadata regarding the planet, and flags concerning the detection of TTVs.

\section{Discussion}

\begin{table}
\centering
\caption{Distribution characteristics for the ephemerides drifts S/N between this work and the previous \exoclock\ publication (first column) and between this work and the ephemerides at the beginning of the project (second column). Planets with TTVs have been excluded. In the ideal case of a normal distribution these parameters should be close to the values in the third column.}
\label{tab:drift}
\begin{tabular}{c c c c}
 						& vs ExoClock II 	& vs initial	& Normal		\\
   						& (180 planets) 	& (450 planets)	& Distribution	\\ \hline
STD						& 1.50				& 37.00			& 1.00			\\ 
Kurtosis				& 1.75				& 262.02		& 0.00			\\ 
68 percentile			& 1.28				& 1.79			& 0.995			\\ 
95 percentile			& 3.42				& 5.23			& 1.959			\\
99 percentile			& 5.13				& 25.96			& 2.576			\\
\end{tabular}
\end{table}

\subsection{Follow-up efficiency}

From the comparison between the ephemerides in \exoclock\ II and this work, we conclude that biases in the ephemerides produced by ExoClock are decreasing. This is based on the fact that the number of significantly improved or drifting ephemerides is very small (Table \ref{tab:before_after}, left column). Moreover, the drifts found between \exoclock\ II and this study are closer to a normal distribution as seen in Table \ref{tab:drift}, first column. These values highlight the reliability of the produced ephemerides and support the view that the ExoClock project is working effectively towards achieving its goal. 

\subsection{Need for continuous monitoring}

As indicated in Table \ref{tab:before_after} (second column), approximately 45\% of the initial ephemerides have large uncertainties or drifts (categories "significantly improved" and "drifting"). This is similar to the percentage reported in \exoclock\ II, indicating that a significant number of ephemerides derived in discovery papers (including TESS discoveries) needs to be corrected to be appropriate for the efficient planning of Ariel. Moreover, as shown in Tables \ref{tab:before_after} and \ref{tab:drift} (first columns), while ExoClock ephemerides have reduced biases, they are not completely bias-free. Our sample of 180 planets in \exoclock\ II is not large enough to determine the coverage needed to produce completely bias-free ephemerides, but we can see that coverages of 60\% or more are necessary to avoid unexpected drifts larger than five minutes in our 2029-predictions. Finally, as discussed in the previous section, some planets show long-term trends. Such trends can only be identified when coverage is close to 100\%. For all these reasons, the effort of follow-up observations is important and continuous monitoring is essential.

The most important factor to increase coverage is to continue integrating all available resources and prioritise accordingly the follow-up observations.

\subsection{Follow-up capabilities}

From the large number of observations obtained so far by the \exoclock\ network and with the TESS light-curves analysed in this study, we can estimate more precisely the capabilities of these resources and plan efficiently for the future. In appendices \ref{sn_exoclock} and \ref{sn_tess} we provide detailed calculation of the signal-to-noise calibration that we performed on both \exoclock\ and TESS data to produce the equations below.

The minimum telescope aperture diameter (in inches) necessary to observe a planet with the \exoclock\ network ($D_{min}$) is given by:
\begin{equation}
D_{min}
=  \frac{0.135 + 10^{-2.99 + 0.2 R}}{5.1  d } \sqrt{\frac{7200 + t_{14}}{900  \pi  t_{14}}}
\end{equation}

\noindent where $R$ is the magnitude of the host star in the $R$ Cousins filter, $d$ is the relative transit depth, and $t_{14}$ is the total duration of the transit in seconds.

\begin{figure}
\centering
\includegraphics[width=\columnwidth]{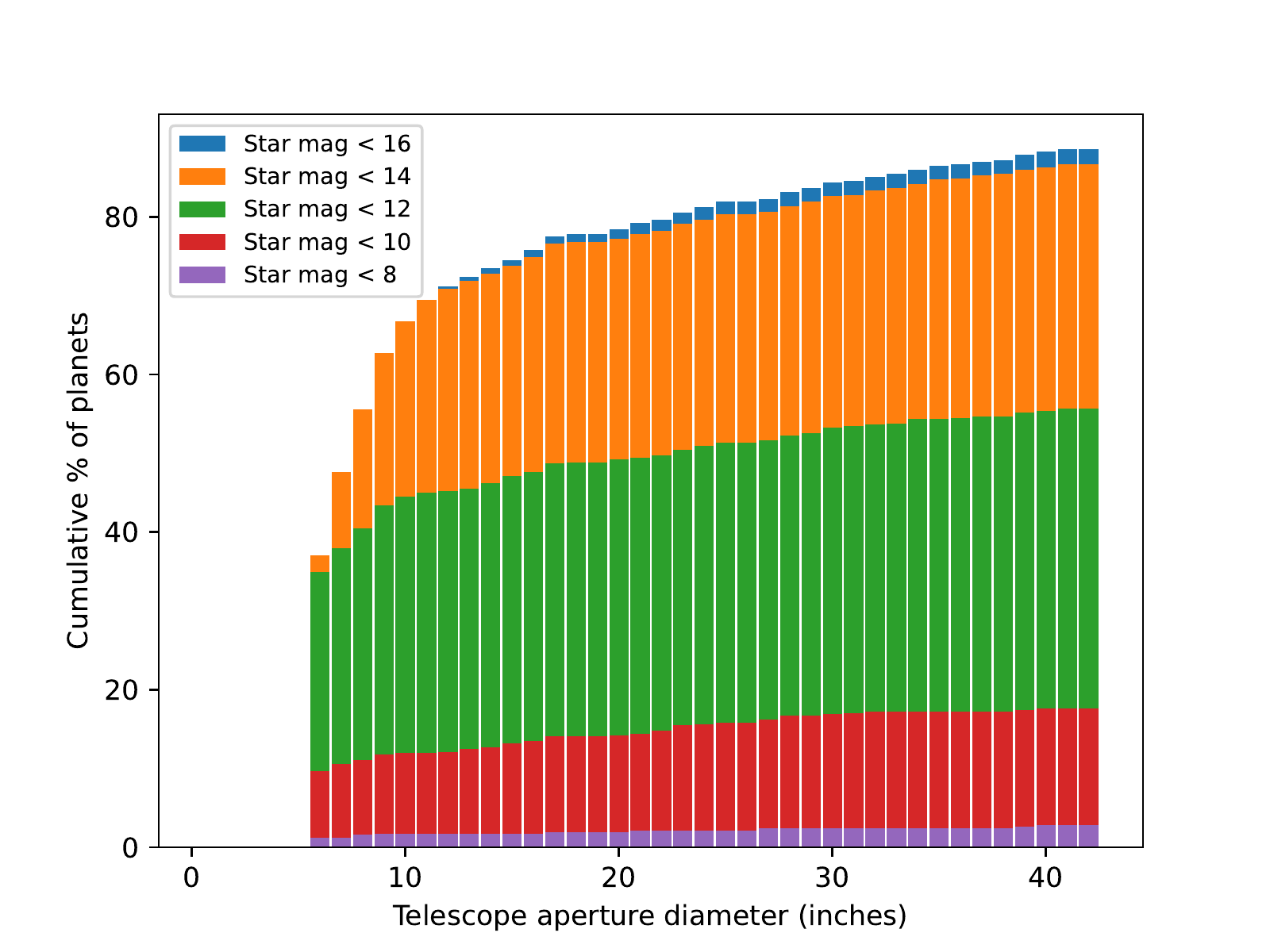}
\caption{Distribution of available planets per magnitude and telescope aperture diameter.}
\label{fig:network_capabilties}
\end{figure}

By placing an upper limit of 40 inches aperture on the \exoclock\ network, we estimate that we can follow-up 88\% of the currently known Ariel candidates. Figure \ref{fig:network_capabilties} shows the distribution of available planets per magnitude and telescope size, where we can see that even with telescopes up to 16 inches, 75\% of the targets can be observed.

For TESS observations, the transit S/N that can be achieved is given :

\begin{equation}
S/N _{transit}^{TESS}
= \frac{0.65 d \sqrt{t_{14}/90} \times 10^{3}}{0.135 + 10^{-2.43 + 0.2  G_{RP} + 0.0039 G_{RP}^2}}
\end{equation}

\noindent where $G_{RP}$ is the magnitude of the host star in the GAIA Rp filter, $d$ is the relative transit depth, and $t_{14}$ is the total duration of the transit in seconds.

By placing a lower limit of S/N = 3 on the TESS observations, we estimate that we can follow-up 90\% of the currently known Ariel candidates. By combining the two resources we can reach up to 95\% of the candidates. For the remaining targets we plan to use other facilities such as CHEOPS \citep{2021ExA....51..109B} and Twinkle \citep{2019ExA....47...29E}, or combined multiple ground-based observations.

With this calibration of our ground-based network and TESS, we are in a good position to achieve the most productive use of both resources. We can avoid wasting valuable space telescope time -- from facilities like CHEOPS or Twinkle -- on following-up targets that can be monitored efficiently from the ground, whilst at the same time we can readily identify the most difficult targets that will definitely require observations from space.

\subsection{TTVs signals}

\begin{figure}
\centering
\includegraphics[width=\columnwidth]{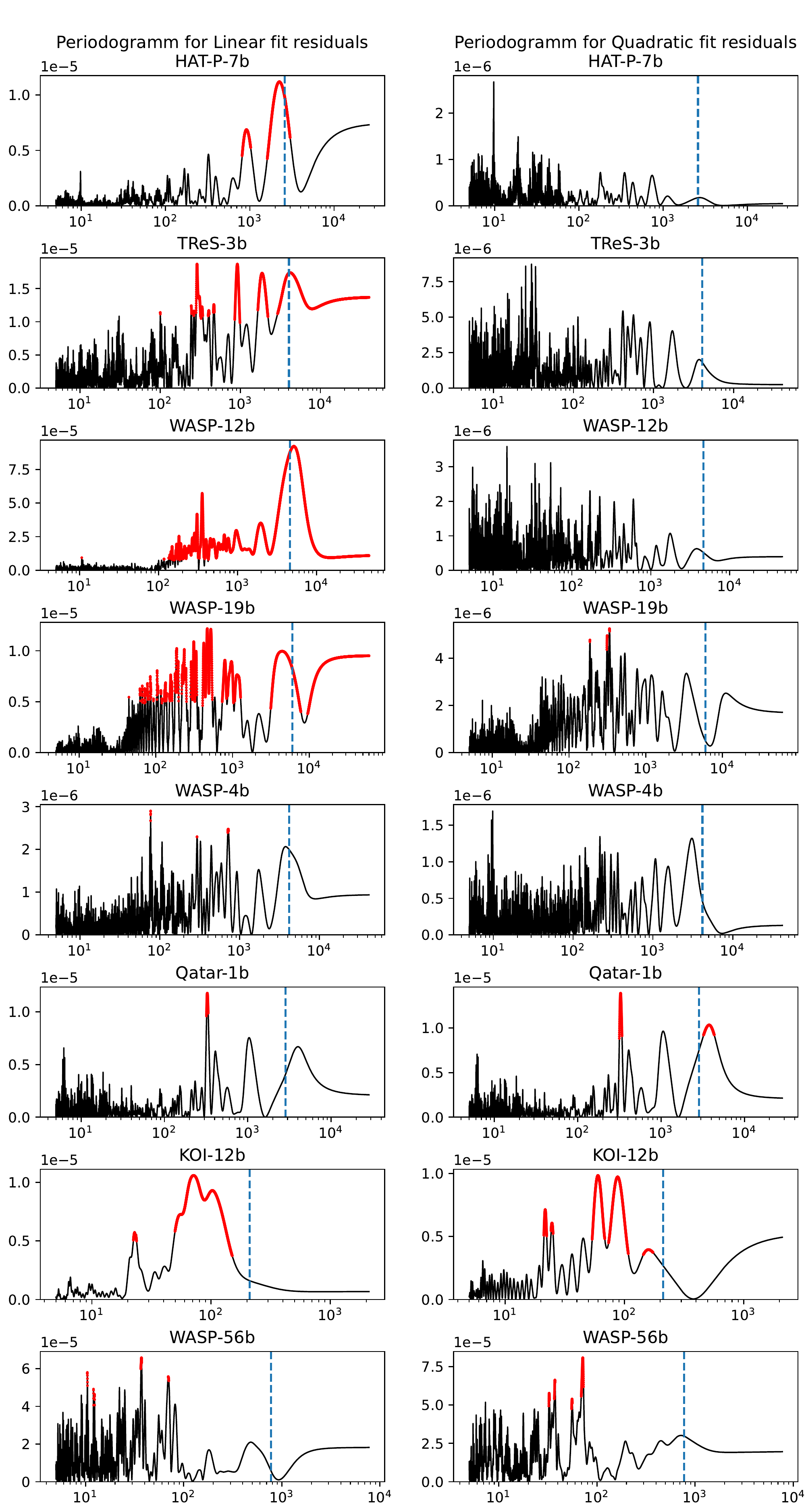}
\caption{Periodogramms for the fitting residuals (linear and quadratic) for the eight planets with TTVs but without transiting companions. The red parts indicate periods with FAP lower than 0.13\% and the vertical line indicates the total time span of the data used.}
\label{fig:periodogramms}
\end{figure}

Our analysis revealed 19 planets with statistically significant signals in the residuals of their linear ephemeris fit (Table \ref{tab:ttvs}). Eleven of these planets -- namely HD106315c, HD108236b, K2-19b, KOI-94c, KOI-94d, KOI-94e, Kepler-18d, Kepler-396c, TOI-216.01, TOI-216.02, TOI-431b -- have one or more additional transiting planets in their planetary systems. Hence, it is no surprise that they show TTVs, due to interaction with other planets in the systems. It is beyond the scope of this work to study the dynamics of these systems, but we are flagging them in the \exoclock\ project, so that observers will continue monitoring them and help with future dynamic analysis. For the remaining eight planets we investigated the different scenarios below. In addition to the periodogramms for the residuals of the linear ephemeris fit, we also applied a quadratic ephemeris fit and studied the periodogramms of these residuals, too (Figure \ref{fig:periodogramms}). 

\paragraph{HAT-P-7b} An attempt to detect a third body in the system, either an additional planet or companion star, has been made using radial velocity data over a two-year span of observations. Analysis from radial velocity data suggest the presence of a companion star but the results were controversial \citep{2009ApJ...703L..99W}. A possible detection of another Saturn-sized planet in the system was also suggested by \cite{2011ApJ...732...41B}, however, the significance for the detection was low. 

\begin{table}
\centering
\caption{Planets identified with deviations from a linear ephemeris. Long-term refers to variations with periodicities that are close to or longer than the total time span of the data used. In brackets we indicate the peak periodicity in epochs (E), while the ``multiple'' label refers to cases were more than one periodicities are significant.}
\label{tab:ttvs}
\begin{tabular}{c c c c}
Planet			& Short-term 		& Long-term 		& Datapoints\\ \hline
HAT-P-7b		& No				& Yes (2243.1 E) 	& 652		\\
HD106315c		& Yes (multiple)	& Yes (multiple) 	& 8			\\
HD108236b		& Yes (multiple)	& Yes (multiple) 	& 9			\\
K2-19b			& Yes (multiple)	& Yes (multiple) 	& 16		\\
KOI-12b			& Yes (71.3 E)		& No 				& 73		\\
KOI-94c			& Yes (15.0 E)		& No 				& 54		\\
KOI-94d			& Yes (13.0 E)		& No 				& 25		\\
KOI-94e			& Yes (7.4 E)		& Yes (38.0 E) 		& 8			\\
Kepler-18d		& Yes (17.6 E)		& No 				& 82		\\
Kepler-396c		& Yes (multiple)	& Yes (44.8 E) 		& 14		\\
Qatar-1b		& Yes (327.0 E)		& No 				& 265		\\
TOI-216.01		& Yes (14.4 E)		& Yes (50.2 E) 		& 15		\\
TOI-216.02		& Yes (35.1 E)		& Yes (111.0 E) 	& 26		\\
TOI-431b		& Yes (multiple)	& No			 	& 21		\\
TrES-3b			& Yes (multiple)	& Yes (4138.1 E)	& 231		\\
WASP-4b			& Yes (multiple)	& No				& 111		\\
WASP-12b		& Yes (multiple)	& Yes (5173.5 E)	& 308		\\
WASP-19b		& Yes (multiple)	& Yes (4465.2 E)	& 119		\\
WASP-56b		& Yes (multiple)	& No				& 25		\\

\end{tabular}
\end{table}

Our results show a significant signal for a long-term periodicity of approximately 2243 epochs, or 4950 days, which is close to, but still lower than, the total time span of the data used. Moreover, we found a significant quadratic term of $6.95\pm0.52 \times 10^{-10}$, and after removing it, the long-term periodicity disappeared. The above results suggest that the signal is not periodic yet, but it has a positive curvature at the moment. This means that the planet is not decaying, leaving the possibilities of a third-body or orbital precession still open.

\paragraph{KOI-12b}

\cite{2017AJ....154...64M} suggested the presence of a second planet based on the same data (Kepler).

Our results indicate a few significant short-term periodicities between approximately 20 and 200 epochs. Moreover, we found a non-significant quadratic term of $-2.49\pm0.97 \times 10^{-7}$, and after removing it, the signals from the short-term periodicities remain strong. From the above we cannot reach a clear conclusion because the multiple short-term periodicities could be caused by stellar activity. More data are required to narrow down the possible scenarios.

\paragraph{Qatar-1b} The first TTVs analysis for the Qatar-1 system was carried out by \cite{2013AnA...555A..92V}. The authors claimed that there are possible TTVs on Qatar-1b either due to a weak pertubator in resonance with Qatar-1b or due to a massive body similar to a brown dwarf. The follow-up TTVs studies by \cite{2015AnA...577A.109M} and \cite{2017AJ....153...78C} did not detect any signal of an additional planet in the system, while \cite{2017NewA...55...39P} found weak evidence of TTVs. It was also reported by \cite{2013AnA...554A..28C} that the orbital period of the planet in the Qatar-1 system is much shorter than the rotation period of the star, so tides produce a decay of the orbit. The most recent analysis, by \cite{2021AJ....161..108S} concluded that no TTV frequencies are identified.

Our data cover a time span that is double compared to previous studies and our results indicate a statistically significant short-term periodicity at approximately 327 epochs or 465 days. Moreover, we found a non-significant quadratic term of $1.13\pm0.72 \times 10^{-10}$, and after removing it, the short-term periodicity was not affected. The above suggest that the signal is periodic, and in combination with the low eccentricity of the planet, this means that a pertubator scenario is favoured.

\paragraph{TrES-3b} So far, studies have concluded that there is no evidence for TTVs for Tres-3b \citep{2013ApJ...764....8K,2017NewA...55...39P}. \cite{2011ApJ...726...94C} mentioned that a long term variability in the light curve of Tres-3b may be due to star spots. Additionally, the lack of periodic TTVs implies that another planetary body is absent, according to the study by \cite{2020AJ....160...47M}. Finally, precession can be ruled out due to the very low value of eccentricity, whereas the possibility of slow orbital decay cannot \citep{2020AJ....160...47M}.

Our results show multiple significant short-term periodicities, as well as one prominent long-term periodicity at approximately 4138 epochs or 5400 days. The long-term signal is longer than the total time span of the data. Moreover, we found a significant quadratic term of $-1.68\pm0.34  \times 10^{-10}$, and after removing it, both the short-term and the long-term periodicities disappeared. The above suggest that the signal is not periodic yet, with a negative curvature at the moment. In combination with the low eccentricity of the planet this means that orbital decay scenario is favoured.

\paragraph{WASP-4b}  From the initial observations of WASP-4b it was assumed that TTVs might be present \citep{2008ApJ...675L.113W}. However, a follow-up study by \cite{2013ApJ...779L..23P} concluded that the system does not show significant TTV trends. \cite{2015MNRAS.450.3101B} proposed that TTVs probably exist in the WASP-4 system with a magnitude of 10-20 seconds and an unknown nature. Additionally, a significant quadratic termm in the O-C diagram was reported in the study by \cite{2019AJ....157..217B} with the most probable explanation being the planets' orbital decay. \cite{2019MNRAS.490.4230S} stated that TTV variations have a smaller magnitude than previously detected, and orbital decay or a third body in the system are both problematic hypotheses. More recently, it was suggested that the line of sight acceleration is the most probable reason for the TTVs \citep{2020ApJ...893L..29B}. Finally, \cite{2020MNRAS.496L..11B} confirmed the existence of quadratic TTVs in the system but without making a new proposal for the origins.

We found a significant quadratic term of $-1.29\pm0.22  \times 10^{-10}$ although a long-term periodicity is not shown. In addition, the short-term periodicities disappeared after removing the quadratic term. The above suggest that timing data alone do not provide any indication towards an interpretation. To further investigate this behaviour, other type of data or a longer time span are required.

\paragraph{WASP-12b} WASP-12b is one of the very first exoplanets with a verified non-linear ephemeris due to orbital decay \citep{2016AnA...588L...6M}. It is also possible that the planet undergoes apsidal precession as the data indicate that the orbit might be slightly eccentric \citep{2020ApJ...888L...5Y}. According to \citep{2017ApJ...849L..11W} the measured rate of the orbital decay would be reasonable only if WASP-12b was a subgiant that experiences evolutionary changes that cause a rapid orbital decay to the planet. TTVs that were reported later support this idea but additional data are needed to confirm this \citep{2018AcA....68..371M}. More recent data concluded that the orbit is decaying with occultation times occurring about four minutes earlier after 10 years \citep{2020ApJ...888L...5Y}. WASP-12b is likely to be engulfed by its host star several million years from now \citep{2020ApJ...888L...5Y}.

Our results show multiple significant short-term periodicities, as well as one prominent long-term periodicity at approximately 5173 epochs or 5650 days. The long-term signal is longer than the total time span of the data. Moreover, we found a significant quadratic term of $-5.24\pm0.17  \times 10^{-10}$, and after removing it, both the short-term and the long-term periodicities disappeared. The above suggest that the signal is not periodic yet, with a negative curvature at the moment. In combination with the low eccentricity of the planet this means that orbital decay scenario is favoured.

\paragraph{WASP-19b}  A non-linear ephemeris was reported previously \citep{2013MNRAS.436....2M,2019MNRAS.482.2065E}. \cite{2020MNRAS.491.1243P} conducted the first empirical study of orbital decay by using 74 complete transit light curves covering a 10 yr period. Their results did not show any sign of orbital decay or periodic variations that could indicate the existence of additional bodies.

Our results show multiple significant short-term periodicities, as well as one prominent long-term periodicity at approximately 4465 epochs or 3520 days. The long-term signal is longer then the total time span of the data. Moreover, we found a significant quadratic term of $-0.87\pm0.13 \times 10^{-10}$, and after removing it, the majority of the short-term periodicities and the long-term periodicity disappeared. The above suggest that the signal is not periodic yet, with a negative curvature at the moment. In combination with the low eccentricity of the planet this means that orbital decay  scenario is favoured.

\paragraph{WASP-56b} 

A search for TTVs in the WASP-56 system was carried out recently in a study by \citep{2021ApJS..255...15W}, but statistically significant trends (at levels of 3$\sigma$) were not found.

Our results indicate a few significant short-term periodicities between approximately 10 and 100 epochs. Moreover, we found a non-significant quadratic term of $-2.09\pm0.82 \times 10^{-8}$, and after removing it, the signals from short-term periodicities above 50 epochs became stronger. From the above we cannot reach a clear conclusion as the multiple short-term periodicities could be caused by stellar activity. With more data in the future we will be able to narrow down the possible scenarios.

\section{Conclusion}

In this study, we present a homogeneous analysis for the ephemerides of 450 planets which are currently known candidates for the Ariel mission. The ephemerides resulted from the integration of data from the \exoclock\ network, mid-time points from the literature, data from the ETD and data from space telescopes (Kepler, K2, and TESS missions). 

The results showed that the ephemerides produced by the \exoclock\ project are less biased and hence more reliable for future predictions compared to the initial ephemerides reported in the literature, while continuous monitoring is necessary, as 40\% of the initial ephemerides for new planets need refinement to achieve the goals of the project. The integrated approach of the project allows us to monitor up to 95\% of the Ariel candidates, while identifying missing data and prioritising observations for specific targets. The \exoclock\ network facilitates effectively the effort of obtaining such high-priority observations, while more difficult targets can be requested to be observed by other space telescopes like CHEOPS and Twinkle.

The \exoclock\ project, after three years of continuous operation, development, and interaction between several communities of academics and non-academics, has became a sustainable platform for providing reliable ephemerides for the Ariel candidate planets. A dynamic evolution of the project is being achieved; new ideas can be implemented with the focus on more specific targets that show special interest (as the ones flagged for TTVs). We plan to continue operating \exoclock\ within the framework of Open Science with the twofold objective of monitoring the ephemerides and fostering the democratisation of science.

\section*{Software and Data} 

Software used: Django, PyLightcurve \citep{Tsiaras2016B2016ascl.soft12018T}, ExoTETHyS \citep{Morello2020}, Astropy \citep{AstropyCollaboration2013}, emcee \citep{ForemanMackey2013}, Matplotlib \citep{Hunter2007}, Numpy \citep{Harris2020}, SciPy \citep{Virtanen2020}.

All the data products and their descriptions can found through the OSF repository with DOI: \href{http://doi.org/10.17605/OSF.IO/P298N}{110.17605/OSF.IO/P298N}.

\section*{Acknowledgements}
The ExoClock project has received funding from the UKSA/STFC grants ST/W00254X/1 and ST/W006960/1. 

This work has made use of data collected with the TESS mission, obtained from the MAST data archive at the Space Telescope Science Institute (STScI). Funding for the TESS mission is provided by the NASA Explorer Program. STScI is operated by the Association of Universities for Research in Astronomy, Inc., under NASA contract NAS 5–26555.

We would like to acknowledge the support provided by the administrators, designers, and developers of the ETD project and of the Czech Astronomical Society both to the ExoClock project but also to the efforts of the whole amateur community through its 10+ years of operation.

This work has made use of observations made by the MicroObservatory which is maintained and operated as an educational service by the Center for Astrophysics, Harvard \& Smithsonian as a project of NASA's Universe of Learning, supported by NASA Award \# NNX16AC65A.

This work has made use of observations made by the LCOGT network, as part of the LCOGT Global Sky Partners project ”ORBYTS: Refining Exoplanet Ephemerides” (PI B. Edwards).

ASTEP has benefited from the support of the French and Italian polar agencies IPEV and PNRA, and from INSU, the European Space Agency (ESA) through the Science Faculty of the European Space Research and Technology Centre (ESTEC), the University of Birmingham, the European Union's Horizon 2020 research and innovation programme (grants agreements n$^{\circ}$ 803193/BEBOP), the Science and Technology Facilities Council (STFC; grant n$^\circ$ ST/S00193X/1), the laboratoire Lagrange (CNRS UMR 7293) and the Universit\'e C\^ote d'Azur through Idex UCAJEDI (ANR-15-IDEX-01).

Members from Silesian University of Technology were responsible for (1) observations planning, (2) automation of observatories work, and (3) processing of data from SUTO network.

A.A. Belinski is supported by the Ministry of Science and Higher Education of the Russian Federation under the grant 075-15-2020-780 (N13.1902.21.0039).

M. Cataneo, E. Russo and M. Cilluffo thank the City Council and Management of Cernusco sul Naviglio for supporting the activity of the Associazione Cernuschese Astrofili and for the construction of the public observatory ``G. Barletta''.

B. Edwards is a Laureate of the Paris Region fellowship program supported by the Ile-de-France Region. This project has received funding under the framework program for research and Horizon 2020 innovation under the Marie Sklodowska-Curie grant agreement no. 607 945298.

P. Gajdoš is supported by the Slovak Research and Development Agency under contract No. APVV-20-0148 and internal grant VVGS-PF-2021-2087 of the Faculty of Science, P. J. Šafárik University in Košice.

C. Haswell and U. Kolb are supported by STFC under grant ST/T000295/1.

M. Mašek is supported by MEYS (Czech Republic) under the project MEYS LTT17006.

L. Mugnai is funded by ASI grant n. 2021-5-HH.0.

{\small
\bibliographystyle{aasjournals}
\bibliography{references,references_manual,references_literature,references_tab_parameters,references_tab_updated_ephemerides}
}

\appendix

\section{SUPPLEMENTARY INFORMATION}

Here we append extra information regarding the data sources and results. More specifically, Table \ref{tab:private_observatories} includes a list with the amateur private observatories contributing to this work, and is followed by a description of the ASTEP telescope. Table \ref{tab:parameters} includes a list with the parameters used in the analysis of individual light-curves and the respective references, where the asterisk indicates orbital parameters ($a/R_s$ or $i$) that were adjusted based on TESS data to match the observed durations.

\begin{longtable}{p{0.38\linewidth} p{0.48\linewidth}}
\caption{Amateur private observatories contributing to this work.}\\
\label{tab:private_observatories}
Observer(s) & Observatory \\ [0.3ex] \hline

Adrian  Jones & I64, Maidenhead, UK \\[-0.5ex]
Leon  Bewersdorff & Observatory Kipshoven, Germany \\[-0.5ex]
Richard  Abraham & The Green Observatory, UK \\[-0.5ex]
Vikrant Kumar Agnihotri & Cepheid Observatory, Rawatbhata, India \\[-0.5ex]
Raniero  Albanesi & 157FrassoSabino \\[-0.5ex]
Enrique  Arce-Mansego & Vallbona Observatory Valencia España \\[-0.5ex]
Matthieu  Bachschmidt & Gonachon, France \\[-0.5ex]
Giorgio  Baj & Observatory- M57, Saltrio Varese (Italy) \\[-0.5ex]
David  Bennett & Rickford Observatory \\[-0.5ex]
Paul  Benni & Acton Sky Portal, Acton, MA, USA \\[-0.5ex]
Patrick  Brandebourg & Observatoire du Guernet, Bretagne, France \\[-0.5ex]
Luboš Brát & ALTAN.Observatory \\[-0.5ex]
Stephen M. Brincat & Flarestar Observatory (MPC:171), San Gwann, Malta \\[-0.5ex]
Sebastien  Brouillard & Observatoire de Saint-Véran - Paul Felenbok - France \\[-0.5ex]
Mauro  Caló & Cavallino Observatory, Tuscany, Italy \\[-0.5ex]
Fran  Campos & Puig d'Agulles Observatory (Vallirana, Spain) \\[-0.5ex]
Alfonso	Carreno & Observatorio Zonalunar MPC-J08 \\[-0.5ex]
Roland  Casali & Alto2000 Observatory, Italy \\[-0.5ex]
Giovanni Battista  Casalnuovo & Filzi School Observatory (Laives - Italy) \\[-0.5ex]
Jean-François  Coliac & OABAC - Observatoire pour l'Astronomie des Binaires et l'Astronomie Collaborative \\[-0.5ex]
Giuseppe  Conzo & Explorer Orbanic Observatory, Croatia \\[-0.5ex]
Mercedes  Correa & Sirius B (Spain) \\[-0.5ex]
Gilles Coulon & Sadr Chili \\[-0.5ex]
Ivan Curtis & YSVP \\[-0.5ex]
Martin Valentine Crow & Burnham Observatory, Burnham on Crouch, UK \\[-0.5ex]
Dominique  Daniel & LMJ-OBS - private observatory - France \\[-0.5ex]
Bruno Dauchet & YSVP (MPC D79) \\[-0.5ex]
Simon Dawes & William James II Observatory, Bexleyheath, England\\[-0.5ex]
Marc  Deldem & Les Barres Observatory, Lamanon, France \\[-0.5ex]
Dimitrios  Deligeorgopoulos & Artemis Observatory, Evrytania, Greece \\[-0.5ex]
Nicolas Esseiva & Observatoire de La Perdrix \\[-0.5ex]
Rafael González Farfán & Uraniborg Observatory, (Écija, Sevilla, Spain) \\[-0.5ex]
Salomon Louw Ferreira & PESCOPE \\[-0.5ex]
Davide  Gabellini & Hypatia Observatory, Italy \\[-0.5ex]
Trevor  Gainey & Kismet observatory, Berkshire, UK \\[-0.5ex]
Josep Gaitan & MAS MOIXA MPC C86 \\[-0.5ex]
Alberto  García-Sánchez & Observatorio Rio Cofio - Robledo de Chavela (Spain) \\[-0.5ex]
Joe Garlitz & Elgin Observatory GJP \\[-0.5ex]
Christophe Gillier & CALA Observatory, France \\[-0.5ex]
Ferran  Grau Horta & Observatori de Ca l'Ou, Sant Martí Sesgueioles, Spain \\[-0.5ex]
Juanjo Gonzales & Cielo Profundo J01 \\[-0.5ex]
Tim  Haymes & Tim Haymes Southside Observatory, N.Oxfordshire, UK \\[-0.5ex]
Ken Hose & Quarryview Observatory / HQR \\[-0.5ex]
Francois  Hurter & Albireo Observatory, Switzerland \\[-0.5ex]
Jens Jacobsen & Egeskov Observatory \\[-0.5ex]
Kevin  Johnson & Holbrook Observatory, East Sussex, UK \\[-0.5ex]
Aziz Ettahar Kaeouach & High Atlas Observatory, Oukaimeden, Morocco \\[-0.5ex]
Bernd Koch & MPC Code B72 \\[-0.5ex]
Didier  Laloum & Observatoire Privé du Mont 40280 Saint-Pierre-du-Mont, France \\[-0.5ex]
Massimiliano  Mannucci & Osservatorio Astronomico Margherita Hack, Firenze, Italy \\[-0.5ex]
Jean-Claude Mario & Observatoire de la cabergue \\[-0.5ex]
Antonio  Marino & Telescopio Remoto Colacevich c/o Osservatorio Astronomico di Capodimonte di Napoli \\[-0.5ex]
Giuseppe  Marino & Osservatorio GAC "Luigi Sturzo", Italy \\[-0.5ex]
Fernando Antonio Martínez-Bravo & Chile \\[-0.5ex]
Paolo Arcangelo  Matassa & P.M.P.H.R. Deep Sky (MPC K81) Atina (FR) Italy \\[-0.5ex]
Philip  Michel & Verulamium Private Observatory, St Albans, UK \\[-0.5ex]
Mike  Miller & Georgetown Observatory, Georgetown, TX USA \\[-0.5ex]
David Molina & AnunakiObservatory, Madrid \\[-0.5ex]
Thomas  Mollier & Tomastro Observatory, France \\[-0.5ex]
Mario  Morales-Aimar & Observatorio de Sencelles, Spain \\[-0.5ex]
Fabio  Mortari & Hypatia Observatory, Italy \\[-0.5ex]
Gabriel Murawski & MGAB Observatory \\[-0.5ex]
Jean-Louis  Naudin & Gatinais French Observatory (GFO) \\[-0.5ex]
Ramon Naves & Montcabrer MPC-213 \\[-0.5ex]
David Néel & Sadr Observatory, Chile \\[-0.5ex]
Alphonso Noschese & Elianto Observatory \\[-0.5ex]
Yenal Öğmen & Green Island Observatory IAU B34 \\[-0.5ex]
Osamu Ohshima & Ohshima Tamashima Observatory \\[-0.5ex]
Zlatko  Orbanic & Explorer Orbanic Observatory, Croatia \\[-0.5ex]
Christian  Pantacchini & Observatoire de BENAYES ; FRANCE \\[-0.5ex]
Nikolaos I. Paschalis & Nunki Observatory, Skiathos, Greece \\[-0.5ex]
Valère  Perroud & Observatoire de Duines, France \\[-0.5ex]
Mark  Phillips & Forthimage Observatory, Edinburgh, Scotland \\[-0.5ex]
Jean-Bernard  Pioppa & La Roque Esclapon - FRANCE \\[-0.5ex]
Jean Plazas & Ribot Observatory \\[-0.5ex]
Jeff  Purcell & Omaha, Nebraska-United States \\[-0.5ex]
Manfred  Raetz & Privat Observatory Herges-Hallenberg, Germany \\[-0.5ex]
François  Regembal & HRT Observatory, Spain \\[-0.5ex]
Jose Angel Carrion Rodrigo & OAO Observatorio Aras de los Olmos \\[-0.5ex]
Lionel  Rousselot & Vierzon Observatory, France \\[-0.5ex]
Xesco  Rubia & Stupa Observatori, Centelles, Catalonia, Spain \\[-0.5ex]
Nello  Ruocco & Osservatorio Astronomico Nastro Verde, Sorrento, Italy \\[-0.5ex]
Mark  Salisbury & POST, UK \\[-0.5ex]
Fabio  Salvaggio & WBRO (K49), Italy \\[-0.5ex]
John  Savage & Z42, Rushay Farm Observatory, Dorset, UK \\[-0.5ex]
Danilo  Sedita & Osservatorio Sedita Castrofilippo, Italy \\[-0.5ex]
Alvaro Fornas Silva & Centro Astronómico Alto Turia (CAAT) \\[-0.5ex]
Nick  Sioulas & NOAK Observatory L02, Greece \\[-0.5ex]
Vojtěch  Školník & Broumov NM Observatory, Czech Republic \\[-0.5ex]
Miroslav Smolka & Motešice Observatory, SK \\[-0.5ex]
Dimitris  Stouraitis & Galileo Observatory, Greece \\[-0.5ex]
Thiam-Guan Tan & Perth Exoplanet Survey Telescope, Australia \\[-0.5ex]
Geoffrey  Thurston & I67, Hartley Wintney, UK \\[-0.5ex]
Fernando Pablo Tifner & MPC I32 \\[-0.5ex]
Andrea  Tomacelli & Telescopio Remoto Colacevich UAN c/o Osservatorio Astronomico di Capodimonte di Napoli \\[-0.5ex]
Alberto  Tomatis & Alto-Observatory, Italy \\[-0.5ex]
Pierre  Valeau & Observatoire de l’Aiguillon sur Mer, France \\[-0.5ex]
Jean-Pascal  Vignes & Deep Sky Chile , Chile \\[-0.5ex]
Alberto	Villa & Oss Astr G.Galilei-Libbiano Mpc code B33 \\[-0.5ex]
Antoni Vives Sureda & Anunnaki observatory \\[-0.5ex]
Kuldip  Vora & Cepheid Observatory, Rawatbhata, India \\[-0.5ex]
Martin Vrašťák & Žilina-Mojš, LSO, Slovakia \\[-0.5ex]
David E. Wright & Yorick Observatory, Hampshire, UK \\[-0.5ex]
Roberto	Zambelli & Roberto Zambelli Observatory \\[-0.5ex]
Martin Zíbar & Chlumčany \\[-0.5ex]

\end{longtable}

ASTEP (Antarctic Search for Transiting ExoPlanets) is a 40 cm telescope installed at the Concordia station, Dome C, Antarctica that operates during the polar winter from March to September \citep{Fressin2005,Daban2010,Mekarnia2016}. The continuous night and excellent atmospheric conditions make it well suited for high precision time series photometry such as exoplanet transit observations. The telescope was installed in 2010 and upgraded in 2022. The project is a collaboration between Laboratoire Lagrange (CNRS UMR 7293), the University of Birmingham, and the European Space Agency.

\begin{longtable}{p{0.15\linewidth} | C{0.2\linewidth} | C{0.2\linewidth} | C{0.1\linewidth} C{0.1\linewidth} C{0.1\linewidth}}
\caption{Parameters used in the analysis of individual light-curves and the respective references, where the asterisk indicates orbital parameters ($a/R_s$ or $i$) that were adjusted based on TESS data to match the observed durations.}\\
\label{tab:parameters}
Planet & Ephemeris (before this update) & Stellar Parameters & \multicolumn{3}{c}{Transit Parameters}  \\ 
       & $T_0$ (BJD$_\mathrm{TDB} - 2450000$) & $T_\mathrm{eff}$ (K) & $R_p/R_s$ & $a/R_s$ & $i$ (deg) \\ 
       & $P$ (days) & $\log(g)$ (cgs)& $e$ & $\omega$ (deg) & \\ [0.3ex] \hline

55Cnce & $2455962.0727_{-0.0007}^{+0.0007}$ & $5234.0_{-30.0}^{+30.0}$ & $0.0187_{-0.0004}^{+0.0004}$ & $3.47_{-0.07}^{+0.07}$ & $83.6_{-0.6}^{+0.6}$ \\ & $0.736545_{-9e-07}^{+9e-07}$ & $4.45_{-0.08}^{+0.08}$ &  -- & -- & \\ & \cite{2019AnA...631A.129S} & \cite{2011AnA...533A.114D} & \multicolumn{3}{C{0.3\linewidth} }{\cite{2019AnA...631A.129S}} \\ \hline
CoRoT-11b & $2454597.6797_{-0.0003}^{+0.0003}$ & $6440.0_{-120.0}^{+120.0}$ & $0.107_{-0.0005}^{+0.0005}$ & $6.89_{-0.08}^{+0.08}$ & $83.17_{-0.15}^{+0.15}$ \\
\hline

\end{longtable}

\newpage

\begin{longtable}{p{0.15\linewidth} | C{0.25\linewidth} | p{0.5\linewidth}}
\caption{Updated ephemerides and data sources.}\\
\label{tab:updated_ephemerides}
Planet & $T_0$ (BJD$_\mathrm{TDB}$) & References for Literature data used  \\
       & $P$ (days) & \\ [0.3ex] \hline

55Cnce & 2459370.807543 $\pm$ 0.000093 0.73654625 $\pm$ 0.00000015 & \cite{2011ApJ...737L..18W} \\ \hline
CoRoT-11b & 2456019.96220 $\pm$ 0.00037 2.99427803 $\pm$ 0.00000049 & \cite{2010AnA...524A..55G} \\ \hline
CoRoT-19b & 2455701.71540 $\pm$ 0.00048 3.8971379 $\pm$ 0.0000016 & \cite{2012AnA...537A.136G} \\ \hline
CoRoT-2b & 2457683.44158 $\pm$ 0.00016 1.74299705 $\pm$ 0.00000015 & \cite{2008AnA...482L..21A}, \cite{2019MNRAS.486.2290O} \\ \hline
EPIC211945201b & 2458094.44793 $\pm$ 0.00024 19.4921498 $\pm$ 0.0000077 &  \\
\hline

\end{longtable}

\section{Transit S/N calculation} 

For a light curve with a standard deviation of $std$, total observing time of $T$, individual points with exposure time of $t_e$, and overheads of $t_o$, the uncertainty  of the relative flux ($\sigma_{F}$) that can be achieved is: 

\begin{equation}
\sigma_{F} = 
\frac{std}{\sqrt{T/(t_e+t_o)}} 
\end{equation}

In the case of a transit (assuming it is square), the transit depth ($d$) is the difference between the out-of-transit relative flux ($F_{oot}$) and the in-transit relative flux ($F_{int}$). Hence the uncertainty on the transit depth ($\sigma_d$) is:

\begin{equation}
\sigma_d
= \sqrt{\sigma_{F_{oot}}^2+\sigma_{F_{int}}^2}
= \sqrt{std^2 \frac{(t_e+t_o)}{T_{oot}} + std^2 \frac{(t_e+t_o)}{T_{int}}}
= std \sqrt{(t_e+t_o)\left(\frac{1}{T_{oot}} + \frac{1}{T_{int}}\right)}
= std \sqrt{\frac{(t_e+t_o) (T_{oot} + T_{int})}{T_{oot}  T_{int}}}
\end{equation}

Hence the square-transit S/N is:

\begin{equation}
\begin{split}
S/N _{square-transit}
= \frac{d}{\sigma_d} 
= \frac{d}{std} \sqrt{\frac{T_{oot} T_{int}}{(t_e+t_o)(T_{oot} + T_{int})}}
\end{split}
\end{equation}

Finally, due to the fact that in reality the transits are not squares and we are also fitting the light curves for extra parameters, there is an additional x-factor to estimate the final transit S/N.

\begin{equation}
S/N _{transit} 
= x S/N _{square-transit}
= x \frac{d}{std} \sqrt{\frac{T_{oot}  T_{int}}{(t_e+t_o) (T_{oot} + T_{int})}}
\label{sn_final}
\end{equation}

From simulations, which we verified with current \exoclock\ and TESS observations, the x-factor is equal to 0.85 for linear or airmass de-trending, and 0.65 for quadratic de-trending. We need to note that for linear and airmass de-trending the x-factor is stable regardless of the length of the out-of-transit observations. However, for quadratic de-trending, in order to maintain the x-factor of 0.65, we need to observe one transit duration before and one after the transit, otherwise the x-factor becomes lower. For example, an observation of a three-hours-long transit with one hour of observations before and after, has an x-factor of 0.5 instead of 0.65.

\section{Transit S/N predictions for ExoClock}  \label{sn_exoclock}

\begin{figure}
\centering
\includegraphics[width=0.49\columnwidth]{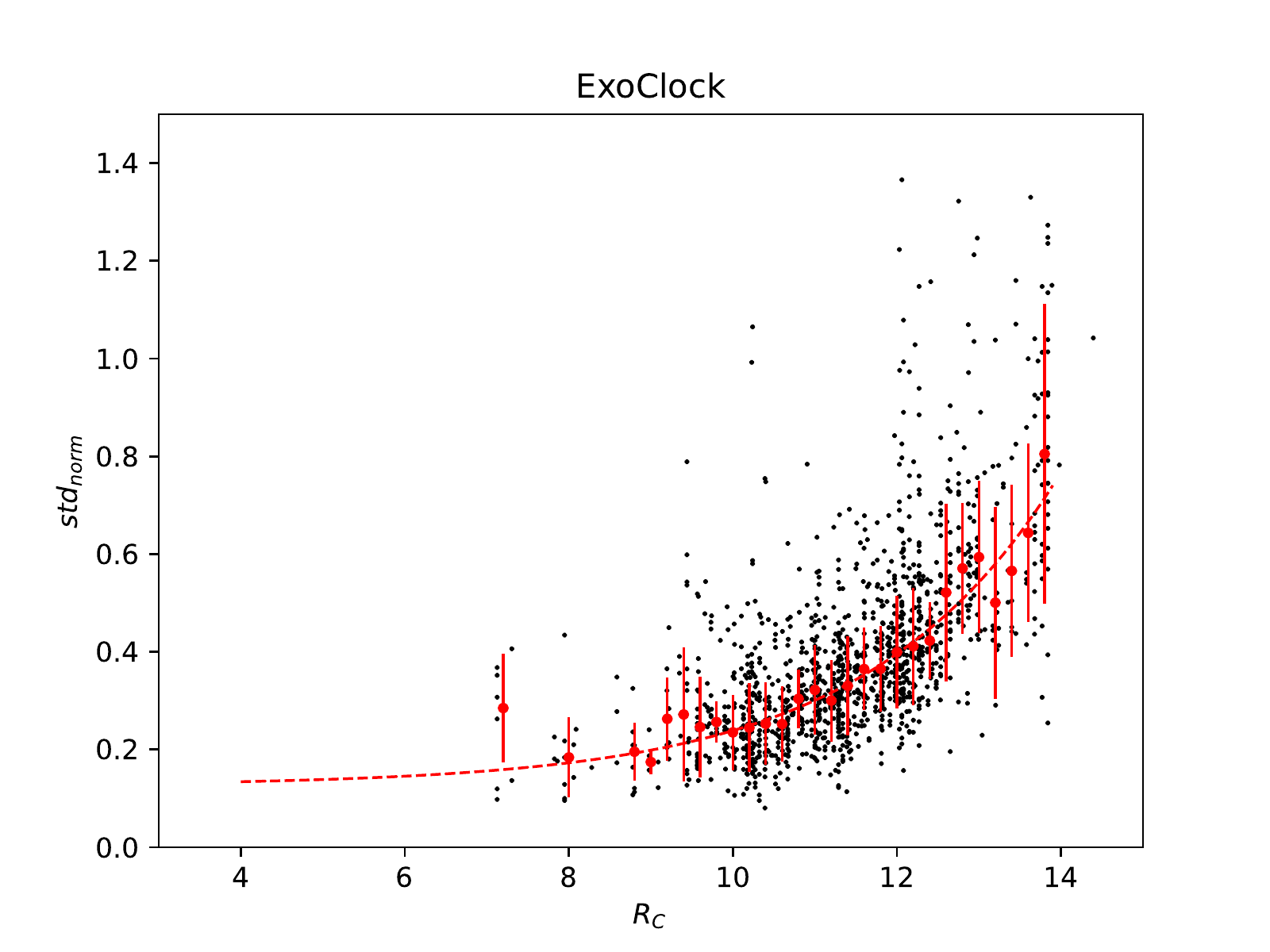}
\includegraphics[width=0.49\columnwidth]{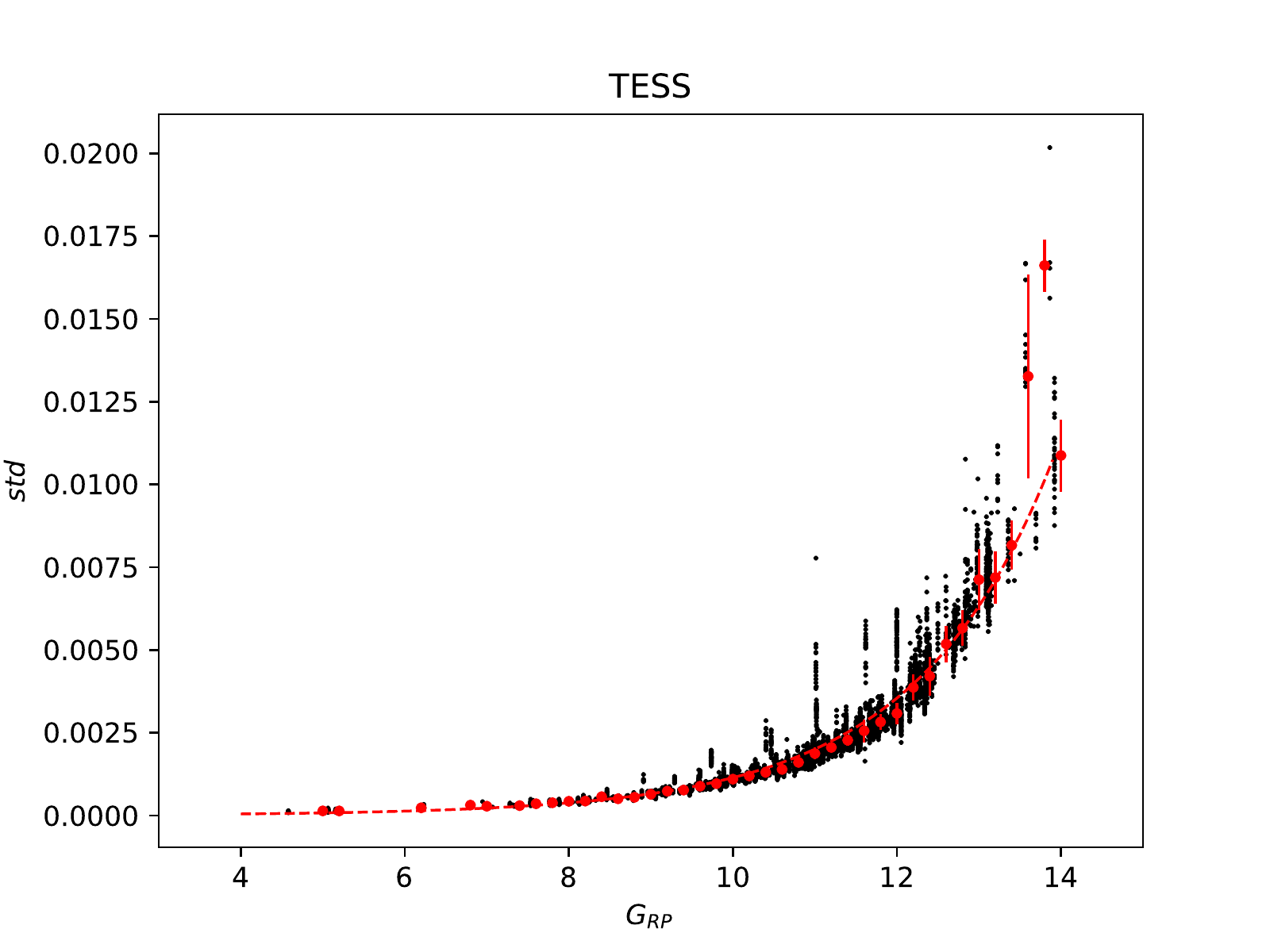}
\caption{Standard deviation of the light-curves as a function of magnitude, for the \exoclock\ (left, normalised for an one-inch telescope and one-second exposure) and the TESS (right) light-curves, together with the models derived. The red errorbars indicate the median and standard deviation of the data in bins of 0.2 magnitudes.}
\label{fig:std_norm_vs_mag}
\end{figure}

To predict the transit S/N we need to have a prediction for all the values included in Equation \ref{sn_final}. The most uncertain one is $std$, which we predicted from the performance of the current telescopes. Figure \ref{fig:std_norm_vs_mag} (left) shows the std of the current observations made using an R Cousins filter, normalised to one second exposure and to a telescope size of one inch, as a function of the $R_C$ magnitude. We have modelled this behaviour as follows:

\begin{equation}
std_{norm}^{ExoClock} = 
0.135 + 10^{-2.99 + 0.2 R}
\end{equation}

Hence, the predicted $std$ for a light curve obtained by a telescope of diameter $D$ and exposure time of $t_e$ will be:

\begin{equation}
std^{ExoClock}
= \frac{std_{norm}^{ExoClock}}{\sqrt{\pi (D/2)^2 t_e}}
= \frac{0.135 + 10^{-2.99 + 0.2 R}}{\sqrt{\pi (D/2)^2 t_e}}
\end{equation}

and the predicted transit S/N will be:

\begin{equation}
S/N _{transit}^{ExoClock}
= x \frac{d \sqrt{\pi (D/2)^2 t_e}}{0.135 + 10^{-2.99 + 0.2 R}} \sqrt{\frac{T_{oot}  T_{int}}{(t_e+t_o) (T_{oot} + T_{int})}}
\end{equation}

The \exoclock\ scheduler calculates the minimum telescope size necessary to observe a transit based on the following assumptions:

\begin{enumerate}
\item the targeted $S/N _{transit}$ is 6
\item the de-trending model is expected to be the airmass model, hence $x$ = 0.85
\item the observation includes one hour before and one hour after the transit, hence $T_{oot}$ = 7200 seconds
\item the observation includes the full transit, hence $T_{int}$ = $t_{14}$ in seconds
\item the overheads and the exposure time are equal, hencee $t_e=t_o$
\end{enumerate}

Hence, the minimum telescope size is:

\begin{equation*}
6
= 0.85 \frac{d \sqrt{\pi (D_{min}/2)^2 t_e}}{0.135 + 10^{-2.99 + 0.2 R}} \sqrt{\frac{7200  t_{14}}{(t_e + t_e)(7200 + t_{14})}}
\end{equation*}
\begin{equation}
D_{min}
=  \frac{0.135 + 10^{-2.99 + 0.2 R}}{5.1  d } \sqrt{\frac{7200 + t_{14}}{900  \pi  t_{14}}}
\end{equation}

\section{Transit S/N predictions for TESS} \label{sn_tess}

As far as the TESS observations are concerned, the calculation is less complicated, as  many of the parameters are fixed. The $std$ can be predicted from the performance of the telescope. Figure \ref{fig:std_norm_vs_mag} (right) shows the $std$ of the current observation, as a function of the $G_{RP}$ magnitude. For TESS, there is no need to normalise to the telescope size and exposure time, as these are fixed. We have modelled this behaviour as follows:

\begin{equation}
std^{TESS} = (0.135 + 10^{-2.43 + 0.2  G_{RP} + 0.0039 G_{RP}^2}) \times 10^{-3}
\end{equation}

Moreover, for TESS the de-trending model is the quadratic ($x$=0.65), the exposure time is two minutes ($t_e=120$) overheads are negligible ($t_o=0$), the observations are continuous, so we can select the out-of-transit observations to be equal to one transit duration before and one transit duration after the transit($T_{oot} = 2t_{14}$ in seconds) and the in-transit observing time is equal to a full transit duration ($T_{int} = t_{14}$ in seconds). Hence, the predicted transit S/N will be:

\begin{equation}
S/N _{transit}^{TESS}
= \frac{0.65 d \sqrt{t_{14}/90}}{0.135 + 10^{-2.43 + 0.2  G_{RP} + 0.0039 G_{RP}^2}} \times 10^{3}
\end{equation}

\end{document}